\documentstyle[amssymb,12pt,aps]{revtex}
\begin{document}
\title{Remark to ''On the Description of Fermion Systems in Boson Representations''}
\author{V.P. Garistov, P. Terziev}
\date{29.11.1998}
\address{Institute of Nuclear Research and Nuclear Energy, Bulgarian\\
Academy of Sciences, Sofia 1784, Bulgaria}
\maketitle

\begin{abstract}
\center{The formulae of boson powers commutation relation used by B.
Sorensen [Nucl. Phys. A {\bf 119}~(No 1) (1968), 65] in his calculations is
erroneous. We provide the correct formulae.}
\end{abstract}

\vspace{1cm}

In the paper of B. Sorensen ''On the Description of Fermion Systems in Boson
Representations'' \cite{1} one can find for bosons powers commutation
relations the formula (2.5 page 66)

\begin{equation}
\left[ a^{n},(a^{\dagger })^{m}\right] =\left\{
\begin{array}{c}
\sum\limits_{l=0}^{n-1}\frac{(l+1)m!}{(m-n+l)!}\;(a^{\dagger
})^{\;m-n+l}a^{l}\;\;\;\text{\ for\ }\;\;n\leq m \\
\\
\sum\limits_{l=0}^{m-1}\frac{(l+1)n!}{(n-m+l)!}\;(a^{\dagger
})^{l}a^{n-m+l}\;\;\;\;\text{\ for\ }\;\;\;n\geq m
\end{array}
\right.
\end{equation}

where $a\;$and$\;a^{\dagger }$\ are\ the annihilation and creation boson
operators.

Dealing with this formula we observed that it is erroneous. Let us for
example compare the straitforwardly calculated results\

\begin{equation}
\left[ a^{4},(a^{\dagger })^{4}\right] =24+96a^{\dagger }a+72a^{\dagger
2}a^{2}+16a^{\dagger 3}a^{3}
\end{equation}

with the corresponding results that gives formula (1)\ ~~for $n=m=4\;:$

\begin{equation}
\left[ a^{4},(a^{\dagger })^{4}\right] =24+48a^{\dagger }a+36a^{\dagger
2}a^{2}+16a^{\dagger 3}a^{3};~
\end{equation}

In this remark, starting with the identity:

\begin{equation}
\left[ A,B^{n}\right] =\sum\limits_{k=0}^{n-1}B^{k}\left[ A,B\right]
B^{n-k-1}
\end{equation}

and putting the operators $\;\;A\longrightarrow a\;;\;B\longrightarrow
a^{\dagger }\;$and then using the induction method, or directly applying the
Leibnitz relations :

\begin{equation}
\begin{array}{l}
\frac{d^{n}}{dx^{n}}(x^{n}\;f)=\sum\limits_{l=0}^{n}%
{n \choose l}%
\left( \frac{d^{n-l}}{dx^{n-l}}x^{n}\right) \left( \frac{d^{l}}{dx^{l}}%
f\right) \\
\text{with}\;\;\;a\longrightarrow \frac{d}{dx}\;\;\;\;\text{and}%
\;\;\;a^{\dagger }\longrightarrow x
\end{array}
\end{equation}

we had derived the right expressions for the boson powers commutation
relation:
\begin{equation}
\left[ a^{n},(a^{\dagger })^{m}\right] =\left\{
\begin{array}{c}
\sum\limits_{l=0}^{n-1}\frac{m!}{(m-n+l)!}%
{n \choose l}%
(a^{\dagger })^{m-n+l}a^{l}\;\;\;\;\;\;\;\text{\ for}\;\;n\leqslant m\  \\
\\
\sum\limits_{l=0}^{m-1}\frac{n!}{(n-m+l)!}%
{m \choose l}%
(a^{\dagger })^{l}a^{n-m+l}\;\;\;\;\;\;\;\;\;\;\text{\ for}\;\;n\geqslant m\
\;\;\;
\end{array}
\right.
\end{equation}

Where $%
{n \choose l}%
\;\;$are\ the\ Binomial\ coefficients.

\end{document}